\documentclass[twocolumn]{webofc}

\usepackage{caption}
\usepackage[varg]{txfonts}   
\begin{document}
\title{Big Bang nucleosynthesis as a probe of new physics}
%
%

\author{\firstname{Carlos A.} \lastname{Bertulani}\inst{1}\fnsep\thanks{\email{carlos.bertulani@tamuc.edu}}  \and
        \firstname{Francis W.} \lastname{Hall}\inst{1}\fnsep\thanks{\email{fhall2@leomail.tamuc.edu}}\and
        \firstname{Benjamin I.} \lastname{Santoyo}\inst{1}\fnsep\thanks{\email{ben@santoyo.us}}
}

\institute{Department of Physics and Astronomy, Texas A\&M University-Commerce, Commerce, TX, United States
          }

\abstract{%
The Big Bang Nucleosynthesis (BBN) model is a cornerstone for  the understanding of the evolution of the  early universe, making seminal predictions that are in outstanding agreement with the present observation of light element abundances in the universe. Perhaps, the only remaining issue to be solved by theory is the so-called “lithium abundance problem". Dedicated experimental efforts to measure the relevant nuclear cross sections used as input of the model have lead to an increased level of accuracy in the prediction of the light element primordial abundances. The rise of indirect experimental techniques during the preceding few decades has permitted the access of reaction information beyond the limitations of direct measurements. New theoretical developments have also opened a fertile ground for tests of physics beyond the standard model of atomic, nuclear, statistics, and particle physics. We review the latest contributions of our group for possible solutions of the lithium problem.
}
\maketitle
\section{Introduction}
\label{intro}
From about 10 seconds to 20 minutes after the Big Bang the universe endured the production of light elements in a process known as Big Bang Nucleosynthesis (BBN) \cite{Gam46,Alp48,abc,Gam48,Alp53,Hoy64,Pee66,Steig07,OlivePDG2010,BK16,Cyb16,Pit18,Fie20,Tur22}. The  temperature during this epoch decreased from about $10^9$ K to $10^8$ K. BBN and its predictions depend strongly on few quantities, namely,  (a) the ratio of  baryon-to-photon number densities, $\eta = n_b/n_\gamma$, assumed to be uniform in space and time during BBN,  (b) the number of neutrinos families $N_\nu$,  and (c) the neutron half-life $\tau_n$.  The Cosmic Microwave Background (CMB) measurement obtained with the Planck satellite yields $\eta = (6.12 \pm 0.06) \times  10^{-10}$ \cite{Kom11}. Precise experiments at the CERN Large Electron–Positron Collider (LEP) have deduced the number of neutrino families  \cite{ALE06} as  $N_\nu = 2.9840 \pm 0.0082$, while neutron lifetime measurements have obtained $\tau_n \simeq 877.75 \pm 28$ s \cite{Pat16}. All these values are in excellent agreement with the BBN model and its predictions of the abundance of light elements. In particular, the BBN model predicts  $\eta=(6.123\pm 0.039)\times 10^{-10}$, compatible with the most recent  measurements of the deuterium primordial abundance \cite{Yeh21}. Nowadays, it is possible to say that the BBN is a parameter-free model, calculated using rather simple computer programs  and yielding consistent predictions for primordial abundances with inputs of experimental nuclear cross sections.
 
\begin{center}
\begin{figure}[t]
\includegraphics[width=8.cm]{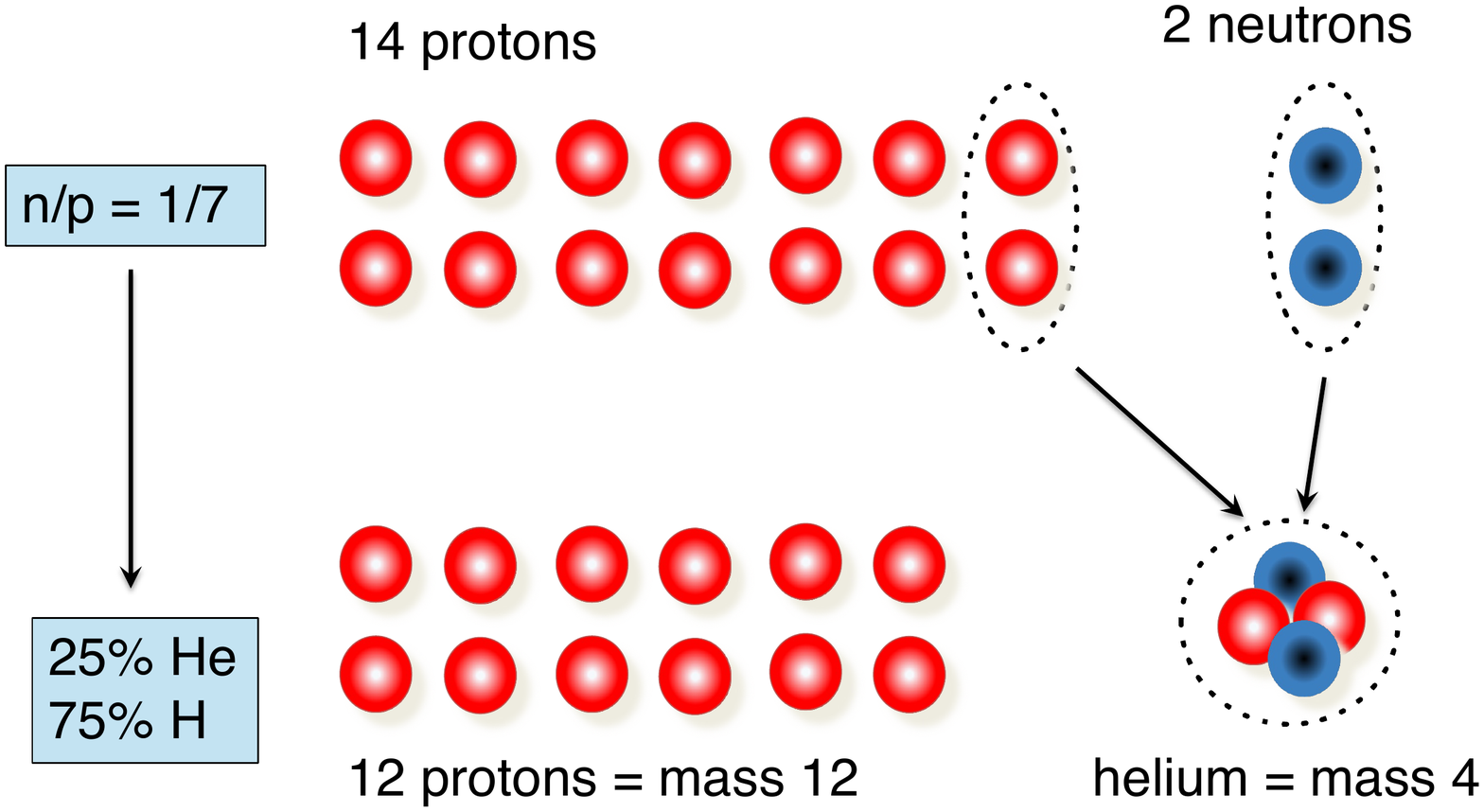}\ \ \ \ \ \ \ 
\caption{{\it Left:} The observed nearly 25\% of helium and 75\% of hydrogen in the universe can be understood if the neutron to proton ratio during the Big Bang nucleosynthesis was equal to ${\rm n/p} = 1/7$. }\label{HeH}
\end{figure}
\end{center}

During the BBN,  the light elements D, $^3$He, $^4$He, and $^7$Li were produced and their abundances have been observed in particular astrophysical environments thought to be remnants of the BBN epoch. The BBN predictions depend on the magnitude of the nuclear cross sections and the nuclear reaction network in a medium where densities and temperatures evolve within the model. One solves the Friedmann's equations for the expansion rate of the universe, 
\begin{equation}
H^2(t) = \left({da/dt\over a}\right)^2 = {8\pi G \over 3} \rho - {k\over a^2} + {\Lambda \over 3},
\end{equation} 
where $H(t)$ is the evolving Hubble constant, $a$ is a universal scale factor for distances, $k$ is the ``curvature'' and $\Lambda$ is the cosmological constant. The universe is assumed to be homogeneous and isotropic (cosmological principle). The Friedmann-Lemaitre-Robertson-Walker (FLRW) metric is adopted to describe distances between neighboring points in the space-time,
\begin{equation}
ds^2 = dt^2-a^2(t)\left[ {dr^2 \over {1-kr^2}}+r^2 \left( d\theta^2 +\sin\theta d\phi^2\right)\right].
\end{equation} 
Additionally, the first law of thermodynamics invokes a relation between density $\rho$ and pressure $p$ in the form
\begin{equation}
{d\rho \over dt} = -3H(\rho + p).
\end{equation} 
Moreover, charge-equilibrium implies that
\begin{equation}
n_b \sum_jZ_jX_j = n_{e^+}-n_{e^-}= \Phi \left({m_e\over T}, \mu_e\right),
\end{equation} 
where $X_i = n_i m_i /\rho$ is the mass fraction of the particle species $i$, with mass $m_i$ and number density $n_i$. Nuclear abundances $Y_i$ are defined so that $X_i=A_iY_i$ is the mass fraction in the environment for a nuclide with charge $Z_i$ and mass number $A_i$, with the normalization $\sum_i A_iY_i = 1$.   The function $\Phi$ incorporates the densities of electrons, positrons and neutrinos. Weak-decays also change the relative numbers of $e^+$ and $e^-$ through $n\leftrightarrow p$ rates.

All nuclear reaction rates include at most four different nuclides and are of the form
\begin{equation}
N_i(^{A_i}Z_i)+N_j(^{A_j}Z_j)\leftrightarrow N_k(^{A_k}Z_k)+N_l(^{A_l}Z_l),
\end{equation}
where $N_m$ is the number of nuclide $m$, with $A_i\geq A_j$ and $A_l \geq A_k$. The abundance changes are given in terms of $dY_i/dt$ which is computed as
\begin{equation}
{dY_i\over dt}=\sum_{j,k,l}N_i\left( - {Y_i^{N_i}Y_j^{N_j}\over  N_i! N_j!} [ij]_k+{Y_l^{N_l}Y_k^{N_k}\over  N_l! N_k!} [lk]_j\right), \label{abund}
\end{equation}
where $[ij]_k$ is the forward reaction rate and $[lk]_j$ is the reverse rate and abundance changes are summed over all forward and reverse reactions involving nuclide $i$. 

The reaction rates in a stellar environment with a temperature $T$ is given by a Maxwell-Boltzmann average over the relative velocity distribution of the pair $[ij]$,
\begin{equation}
[ij]_k=\langle \sigma v \rangle_{[ij]_k} = \sqrt{8 \over \pi \mu (kT)^3} \int_0^\infty \sigma(E) E \exp\left( -{E\over kT}\right) dE, \label{avemb}
\end{equation}
where $\mu$ is the reduced mass of $[ij]$, $N_A$ is Avogadro's number,  $k$ is the Boltzmann constant, and the reaction cross section and its energy dependence is denoted by $\sigma(E)$.

Using  $k=0$ and $\Lambda=0$ for the epoch right after the baryogenensis, the standard BBN model described above leads to the formation of deuterons  about $\sim 3 $ minutes after the Big Bang. The  p(n,$\gamma$)d reaction flow for deuteron formation is strongly dependent on the  value of $\eta$.  Deuterons were immediately destroyed with the creation of $^3$He nuclei by means of d(p,$\gamma)^3$He and d(d,n)$^3$He reactions and the formation of tritium  with the d(d,p)t reaction. Subsequently, $^4$He were formed via the $^3$He(d,p)$^4$He and t(d,n)$^4$He reactions. This chain of reactions yields a universe with about 75\% of hydrogen and 25\% of helium. As the universe expanded and cooled down, nucleosynthesis continued at a smaller rate, leading to very small amounts of  $^7$Li and $^6$Li. Heavier elements such as carbon, nitrogen or oxygen were created in very tiny amounts, of the order of $< 10^{16}$ relative abundance \cite{Coc14}. 

The prediction of 75\% of hydrogen and 25\% of helium mass fraction in the universe is one of the major accomplishments of the Big Bang model \cite{Steig07}. It solely relies on the the neutron-to-proton ratio being equal to n/p = 1/7 at the BBN epoch, as schematically shown in Fig. \ref{HeH}. The n/p = 1/7 value is obtained by solving the Big Bang standard model (see Figure 1 of Ref. \cite{Steig07}).

\begin{center}
\begin{figure}[t]
\includegraphics[width=8.cm]{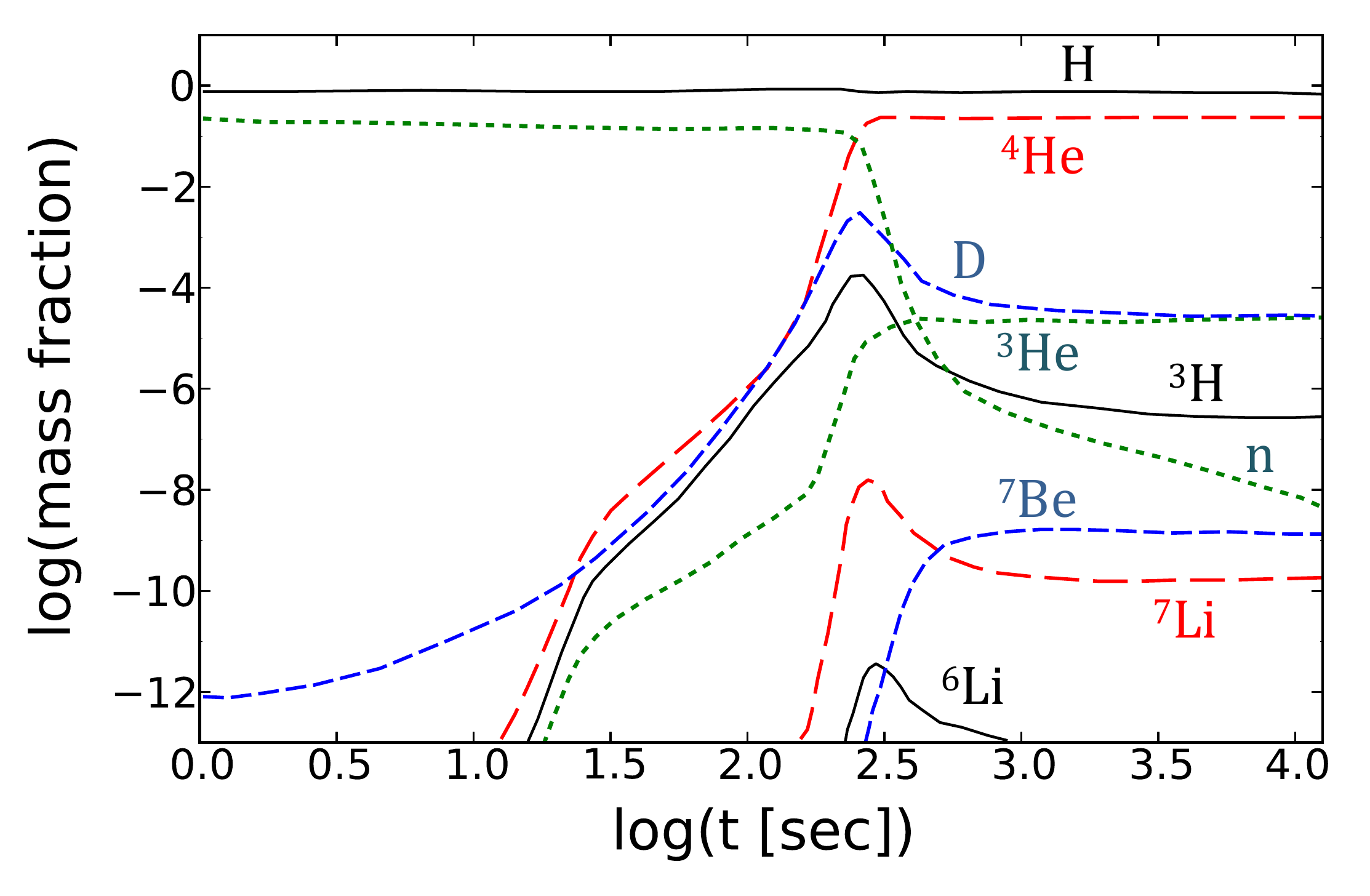}\ \ \ \ \ \ \ 
\caption{Mass (for $^4$He) and number (for other elements) fractions of nuclei in the universe as a function of time.}\label{bbn}
\end{figure}
\end{center}

The most relevant BBN reactions are listed in Table \ref{rent}.  The predicted mass and number fractions of light elements produced during the BBN are shown in Fig. \ref{bbn} as a function of time. One notices that elements up to $^7$Be and $^7$Li are predicted in reasonable amounts. But $^7$Be decays by electron capture in about 53 days. Therefore, all $^7$Be produced during the BBN is observed today as $^7$Li.

\begin{center}
\begin{table}[htbp]
\vspace{0.0cm}
\begin{center}
\scalebox{0.9}{\begin{tabular}{|l|l|l|l|l|}
\hline
\hline
${\rm n} \leftrightarrow {\rm p}$  &${\rm p(n,}\gamma{\rm )d}$  & ${\rm d(p,}\gamma)^3{\rm He}$ & ${\rm d(d, p)t}	$ \\ \hline\hline
${\rm  d(d, n)}^3{\rm He}	$ &$ ^3{\rm He(n,p)t}$  &${\rm  t(d,n)}^4{\rm He}$ & $^3{\rm He(d, p)}^4{\rm He}  $  \\ \hline
$^3{\rm He}(\alpha,\gamma)^7{\rm Be}$  & ${\rm t(}\alpha,\gamma)^7{\rm Li}$ & $^7{\rm Be(n,p)}^7{\rm Li}$  &$^7{\rm Li(p,} \alpha )^4{\rm He}	$ \\ \hline
\hline
\end{tabular} }
\end{center} 
\caption{Most relevant BBN reactions. }\label{rent} 
\end{table} 
\end{center}

BBN predicts  a $^7$Li/H of abundance of $\sim 10^{-10}$ and a $^6$Li/H abundance of $\sim 10^{-14}$.  After the BBN era ($\gtrsim 20$ min), the universe cooled down and no more elements were formed until the first stars were born, about 100 million years later.  After the BBN and substantial formation of stars,  $^6$Li and $^7$Li could  be produced in spallation processes by cosmic rays formed by star ejecta. $^7$Li can also be synthesized in novae or  during AGB stars pulsations.  Astronomical observation suggest that the $^7$Li abundance does not depend on ``metallicity" (the content of elements heavier than $^4$He) in metal-poor stars. These stars have small Fe/H abundances and are observed in the galaxy halo.  The $^7$Li abundance in such stars is nearly constant,  a phenomenon known as the ``Spite plateau'' \cite{Spi82}. 

Low metallicity stars with masses $M< M_\odot$ and life expectancies greater than the age of the universe have (in principle) no convective zone near their surfaces. Lithium present in their surfaces cannot be brought to deeper zones and be destroyed where temperatures exceed 1 GK. The depth of the convection zone decreases with the surface temperature, and that is the reason why astronomers select stars with $T_{eff} > 6000 $ K (warm) for such observations (for more details, see Ref. \cite{FO22}).
Other observations of low-metallicity stars apparently contradict the inferences based on the Spite plateau \cite{Aok09,Sbo10}. Using a variety of observations and laboratory data on nuclear reactions, and  $\eta = (6.07 \pm 0.07) \times 10^{-10}$ \cite{Kom11}, the BBN model predicts a $^7$Li abundance of ${\rm Li/H} = (4.16 - 5.34) \times 10^{-10}$ \cite{Piz14}
 whereas observations from metal-poor halo stars obtains ${\rm Li/H}  = (1.58+ 0.35-0.28) \times 10^{-10}$ \cite{Sbo10,Coc12}.  This value is about 3 times smaller than the one predicted by the BBN model and is the source of the so-called ``lithium puzzle''.

For some time, it was apparent that there was a second lithium puzzle, involving the BBN ${}^{6}{\rm Li}$  abundance via the $^{2}{\rm H}(\alpha,\gamma)^{6}{\rm Li}$ reaction. $^6$Li (and $^7$Li) is also formed  in  cosmic ray spallation reactions in the interstellar medium out of which (halo) stars are born. The second lithium puzzle concerns the BBN predictions for the isotopic ratio  ${}^{6}{\rm Li}/{}^{7}{\rm Li}  \sim  10^{-5}$ \cite{Piz14,MSB16} being smaller than the observations  $^{6}{\rm Li}/{}^{7}{\rm Li}  \sim 5 \times 10^{-2}$ \cite{Asp06}.  This puzzle is controversial due to the complexities of 3-dimension calculations with convection and non-local thermodynamical equilibrium in the photo-sphere of metal-poor stars. These complexities weaken the possibility of a second lithium problem for the  $^{6}{\rm Li}/{}^{7}{\rm Li}$ isotopic ratio. Recent observations   seem to indicate that the second cosmological lithium problem does not exist, based on the lack of evidence of $^6$Li in Spite Plateau stars \cite{LMA13,Wang2021}.  

\section{Nuclear and atomic physics}
\label{sec-2}
As mentioned above,  the most relevant BBN nuclear reactions are shown in Table 1. The reaction network leads a significant production of D, $^3$H, $^3$He, $^4$He, $^6$Li, $^7$Li and $^7$Be and to the production of small traces of carbon, nitrogen and oxygen, at the $10^{-15}-10^{-25}$ abundance yield. It is nearly useless to make predictions beyond such nuclei because they are overwhelmingly produced in stars although the inclusion of standard nuclear reactions networks,  such as the famous CNO cycle have been carried out for clarity  \cite{Coc12}.  

Our calculations presented in Fig \ref{bbn} for the BBN abundances of H, D, $^3$H, $^3$He, $^4$He, $^6$Li, $^7$Li and $^7$Be as a function of time  \cite{BK16} have been performed with an extended version of a code  originally due to  Wagoner \cite{Wag69}, reformulated by Kawano \cite{Kaw92} and later modified to include the most recent updated nuclear reaction rates \cite{BHS2022} and  new physics aspects discussed in this short review. In Fig. \ref{bbn} the $^4$He mass fraction is represented by the dashed red curve, the deuterium abundance ratio D/H by the blue dashed curve, the $^3$He /H abundance by the dotted green curve, the $^3$H/H abundance by the solid black curve, the $^7$Be/H abundance by a  blue dashed curve, and  the $^7$Li/H abundance by a red dashed curve. New experimental measurements of nuclear reaction cross sections, including those  reported with the Trojan Horse Method (THM) \cite{Piz14}, have been incorporated in our compilation using R-matrix fits \cite{BHS2022} .   

\subsection{Electron screening and clustering effects}

\begin{center}
\begin{figure}[t]
\includegraphics[width=8.cm]{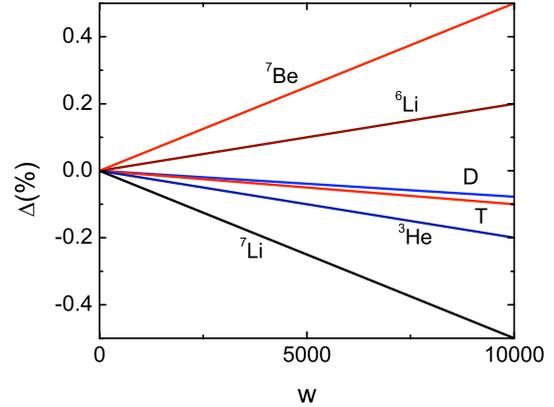}\ \ \ \ \ \ \ 
\caption{Variation (in percent) of the abundances of  light nuclei produced in the BBN as a function of an artificial  multiplicative factor $w$  enhancing Salpeter's formulation of screening \cite{Wan11}.}\label{bbnscr}
\end{figure}
\end{center}

One of the problems in deducing the reaction cross sections at the low astrophysical energies needed for the BBN is the cross section enhancement due to the electrons in the atomic targets in laboratory experiments. Also, within in the stellar plasma, free electrons are further responsible for enhancements of the reaction rates. The reaction rate enhancement in the plasma can be approximated rather well by the factor 
\begin{equation} 
f(E) = {\sigma _{s}(E)\over \sigma _{b}(E)},
\end{equation} 
where $\sigma _{s}$ denotes the screened and $\sigma _{b}$ the non-screened, or ``bare'' fusion cross section. Using the Debye-H\"uckel approximation the screened Coulomb potential of a nuclear species $i$ can be written as  \cite{Sal54} 
\begin{equation}
V_i(r) = {e^2 Z_i \over  r} \exp\left(-{r\over R_D} \right).
\end{equation} 
The Debye radius entering this equation is  related to the ion number density $n_i$ and all environment charges $j$ as  
\begin{equation}
R_D = {1 \over  \zeta}\left( {k T \over 4 \pi e^2 n_i} \right)^{1/2},
\end{equation} where 
\begin{equation} \zeta = \left[ \sum\limits_j X_j {Z_j^2\over A_j} +  \chi \sum\limits_j X_j {Z_j\over A_j} \right]^{1/2}.
\end{equation} 
Here, $X_j$ denotes the mass fraction of particle $j$, $T_6$  the temperature  in units of 10$^6$ K and $\chi$ is a correction due to effects of electron degeneracy. During the BBN, electrons and positrons  are produced via the reaction $\gamma \gamma \rightarrow e^+e^-$. As the universe cooled down, the number density of electrons decreased strongly with $T_6$, from about $10^{32}$/cm$^3$ at $10^{10}$ K to $10^{20}$/cm$^3$ K at $3\times 10^8$ K  \cite{Wan11}. For comparison,  the electron density in the core of the sun ($T=15 \times 10^6$ k) is $n_e^{sun}\sim 10^{26}$/cm$^3$.   But the baryon number density is much lower during the BBN than in the core of the sun and, due to charge neutrality, the number of excess electrons during the BBN is roughly equal to the number of protons.  These excess electrons are the most relevant for the screening effect.  

An attempt was reported in Ref. \cite{Wan11} to understand the effects of electron screening in the BBN production of light elements finding that these effects are negligible. A fudge factor $w$ artificially reducing the Debye radius $R_D$  was introduced to simulate any unusual enhancement of the electron screening in nuclear reactions during the BBN. The results for the modified light element abundances are shown in Fig. \ref{bbnscr}. The abundances are very resilient to a modification of the electron screening even if an extremely small Debye radius is assumed. Later, additional works have extended this study to include relativistic corrections \cite{Fam16,Luo20}. 

It became clear that electron screening  cannot be responsible for the deficiency in the observed lithium abundance. Another publication studied the clustering in light nuclei and their influence in nuclear reactions at the BBN relevant energies. Clustering could perhaps explain part of the discrepancies of the experimental extraction of the electron screening enhancement factor $f(E)$ and the predictions of theory \cite{Spit16,BS17}.

\begin{center}
\begin{figure}[t]
  \centerline{ \includegraphics[width=8.cm]{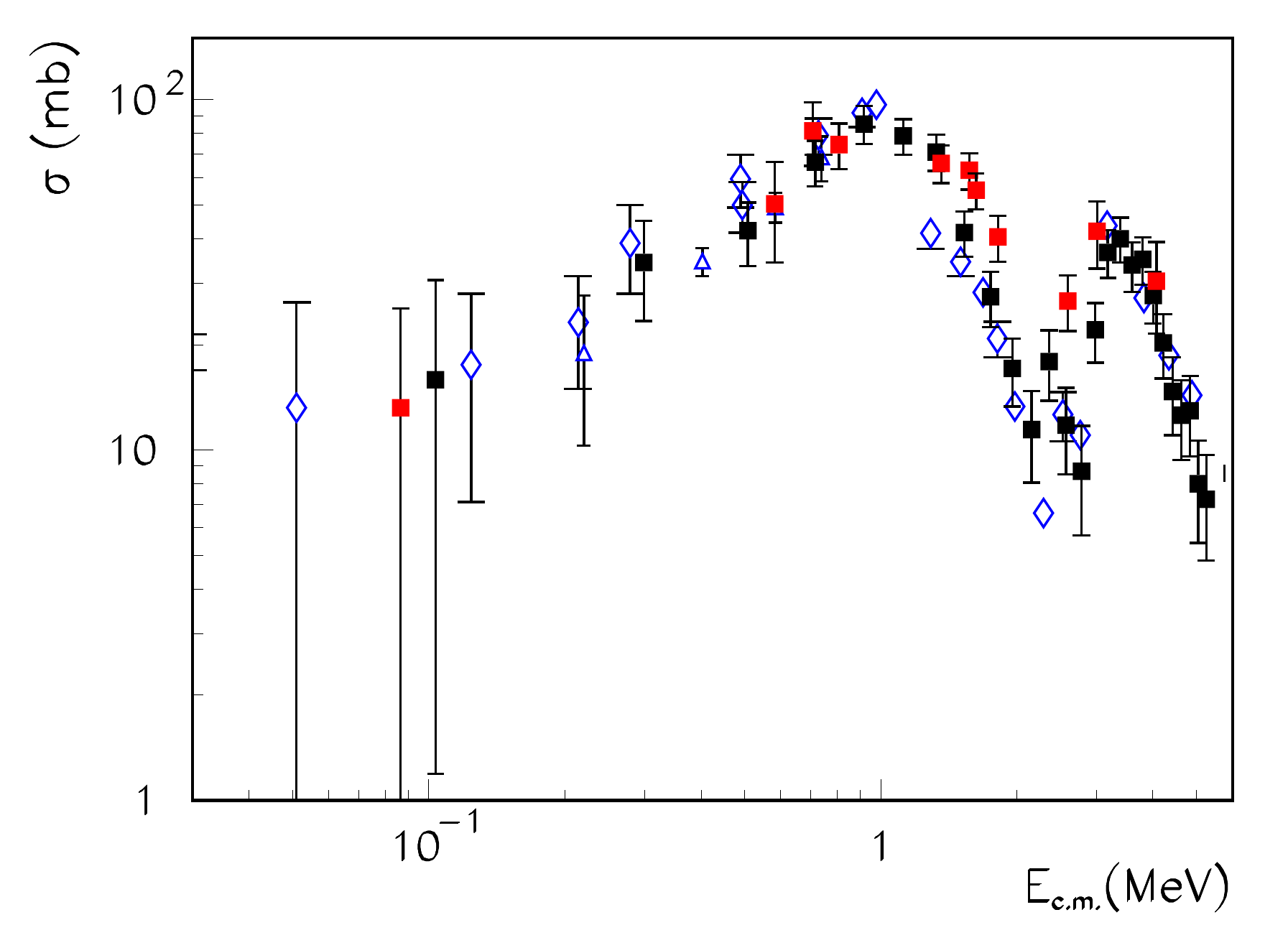}}
\caption{$^7$Be(n,$\alpha$) cross section obtained with  the THM using the  $^7$Li(p,$\alpha)^4$He mirror reaction
(full red squares) and  $^3$He THM breakup data (full black squares) \cite{Lam17}. Other data compiled in Ref. \cite{Hou15} are displayed as empty blue diamonds. Additionally, data from  Ref. \cite{Kaw17} are displayed as empty blue triangles. }\label{fig1}
\end{figure}
\end{center}

\subsection{Nuclear reaction rates}

In Table \ref{tabbbn} we compare our BBN calculations with observations. $Y_p$ is a historical denomination of the $^4$He mass fraction. Its value is taken from Ref. \cite{Aver21}.  The deuterium abundance  ${\rm D/H} = (2.527 \pm 0.03) \times 10^{-5}$ adopted from Ref. \cite{RJ17,Coo18} is in accordance with the value $100\Omega_b h^2 \ ({\rm BBN}) = 2.225 \pm 0.016$ obtained from the observations of the cosmic microwave background \cite{Kom11}. Here, $\Omega_b h^2$ is the baryon density,  where $h$ is the reduced Hubble constant \cite{Fie20}. The $^3$He/H abundance is extracted from Ref. \cite{BRB02}. Finally,   the lithium abundance  is adopted from Ref. \cite{Sbo10}. It is in visible disagreement with astronomical observations (${\rm BBN/obs} \sim 3$). 

Numerous attempts have been discussed in the literature as possible ways to explain the discrepancy between the observed and the calculated $^7$Li abundance. A few attempts concentrated on the  $^7$Be abundance because $^7$Be could be destroyed during the BBN. $^7$Be decays  in about $53.22 \pm 0.06$ days by electron capture to the $^7$Li ground state and also to the first excited (0.477 MeV)  of $^7$Li. Thus,  $^7$Be  created during the BBN will count towards the $^7$Li primordial abundance.  $^7$Be could be destroyed by (n,p) and/or (n,$\alpha$) reactions and explain the smaller lithium abundance. Refs. \cite{Brog12,Hou15,Bar16,Kaw17,Lam17} have experimentally investigated this possibility. For example, Lamia et al. \cite{Lam17} measured the  $^7$Be(n,$\alpha$) reaction cross section using the Trojan Home Method (THM)  for the  $^7$Li(p,$\alpha)^4$He mirror reaction and corrected it to account for Coulomb effects, as shown in Figure \ref{fig1}. The cross section for $^7$Be(n,$\alpha$) obtained with the THM is located within the Gamow window at the BBN temperatures and leads to a reaction rate smaller by a factor $\sim 10$ to the one adopted by Wagoner \cite{Wag69}. A BBN calculation using the new reaction rate results in a $^7$Li/H abundance of $2.845 \times 10^{-11}$ and a $^7$Be/H abundance of $4.156 \times 10^{-10}$. This modification yields a BBN lithium/H abundance of $4.441 \times 10^{-10}$. No substantial modification of the BBN abundances prediction for other light nuclei is confirmed. 

A theoretical study of the effect of $^7$Be($\alpha,\gamma$) on $^7$Be destruction was further carried out in Ref. \cite{Har18}, implying that the $^7$Be abundance would only be changed appreciably  if an unexpectedly  strong resonance would exist very close to energy threshold for $\gamma$ emission. At present, it appears that this resonant state has not been identified experimentally.

\begin{table}[htbp]
\vspace{0.0cm}
\begin{tabular}{|c|c|c|c|c|c|c|c|}
\hline
\hline
Yields &   Calculation   & Observation\\  \hline
$Y_p$ &\text{0.2485{\footnotesize +0.001-0.002}}&$0.2453 \pm 0.0034$ \\ \hline
D/H ($\times 10^{-5}$)&\text{2.692{\footnotesize +0.177-0.070}}&$2.527 \pm 0.03$ \\ \hline
${^3}$He/H ($\times 10^{-6}$) &\text{9.441{\footnotesize +0.511-0.466}}&$\geq 11.\pm 2.$\\ \hline
${^7}$Li/H ($\times 10^{-10}$)&\text{4.283 {\footnotesize +0.335-0.292}}&$\text{1.58{\footnotesize +0.35 -0.28}} $ \\ \hline
\hline
\end{tabular} 
\vspace{0.0cm}
\caption{\label{tab:table2} Standard BBN yields of light elements updated with input of recent experimental  data for nuclear reactions. The table includes results based on astronomical observations. The mass fraction for $^4$He is historically denoted by $Y_p$. It is taken from Ref. \cite{Aver21}.  The  deuterium abundance  ${\rm D/H} = (2.527 \pm 0.03) \times 10^{-5}$ \cite{RJ17,Coo18} is in accordance with $100\Omega_b h^2 \ ({\rm BBN}) = 2.225 \pm 0.016$ deduced from the observations of the cosmic microwave background \cite{Kom11}. The  $^3$He abundance is taken  from Ref. \cite{BRB02},  whereas the lithium abundance  is from Ref. \cite{Sbo10}. 
}
\label{tabbbn}
\end{table}     

Another attempt to explain the lithium puzzle was reported using modifications of the nuclear reactions for the  $^7$Li(d,n)2$^4$He reaction rates \cite{Hou21}. Apparent uncertainties still exist  for this reaction used  in BBN calculations. The latest data for the $^7$Li(d,n)2$^4$He reaction rate focuses on three near-threshold $^9$Be excited states. The contribution from a subthreshold resonance at 16.671 MeV in $^9$Be was accounted for in Ref. \cite{Hou21}. The reaction rate for $^7$Li(d,n)2$^4$He rate was found to become smaller than previous estimations by about a factor of 60 at typical temperatures considered during primordial nucleosynthesis. The new rates are shown to have a small impact on the final light element abundances in the standard BBN model. Only variations of the order of 0.002\% compared with other calculations were found. But, for nonuniform density BBN models, the final $^7$Li abundance increases up to 10\% while those for other light nuclides with mass number $A >7$ can increase by about 40\%  \cite{Hou21}. Thus,  the cosmological lithium problem seems to be resilient  from the standpoint of nuclear physics.

BBN predicts an isotopic ratio of $^6{\rm Li}/^7{\rm Li} \sim 10^{-5}$, whereas  observation seemed to yield $^6{\rm Li}/^7{\rm Li} \sim 5 \times 10^{-2}$ \cite{Asp06}. Due to the apparent puzzle, a re-analysis of the reaction $^4$He$({\rm d},\gamma)^6$Li was made in  Ref. \cite{MSB16}, leading to new predictions for the angular distributions of the gamma-ray. A two-body potential model was adopted to calculate the cross section for this reaction at the BBN relevant energies \cite{radcap,MSB16}. The potential parameters were fixed to reproduce experimental phase shifts and the measured ANCs reported in the literature.  A good agreement with experimental data of the LUNA collaboration \cite{LUNA} was found.  This work is again consistent with BBN predictions for the lithium isotopic ratio yielding  $^6{\rm Li}/^7{\rm Li} = (1.5 \pm 0.3) \times 10^{-5}$. However, the non-detection of $^6$Li in metal poor stars by the EXPRESSO collaboration \cite{Wang2021} implies a simpler solution to the second cosmological lithium problem. According to Ref. \cite{Wang2021} this problem never existed, because there is no evidence of $^6$Li in Spite Plateau stars.

As for the puzzle concerning the $^7$Li abundance, it does not seem possible that it could be solved by accurate laboratory experiments  to measure nuclear reaction cross sections. Advanced modeling of nuclear astrophysical reactions also do not seem to solve the problem. But a considerable number of theoretical works to elucidate the lithium puzzle resort to ideas outside the range of nuclear physics. Perhaps, many physics processes could be different now than 13.8 billions years ago. Particles that could be decoupled from known ones, other sort of interactions, modifications in the values of fundamental constants, the possibility of non-standard BBN models, and other intriguing ideas have been explored and reported in the literature. Next we discuss a few ideas developed and tested by our group.       
 
\section{Dark parallel universes}

One of the biggest puzzles in present day cosmology are the observations that most of the matter in the universe is composed of an obscure kind of matter, denoted as Dark Matter (DM). This DM seems to interact very weakly with the visible universe, mostly through gravitation. Astronomical observations of galaxy clusters dynamics and of anisotropies of the Cosmic Microwave Background (CMB) are the main proofs of the DM existence. Intensive experimental searches are being carried out to determine how DM might interact with visible matter other than by gravitation
 \cite{Feng2010,bertone05,bertone10}.   DM could be formed by some kind of  Weakly Interacting Massive Particles (WIMPs), it could be composed of supersymmetric particles, or maybe some sort of sterile neutrinos, or even by other unknown particles. 
 
 Hypothetically,   DM could lie in a separate mirror sector of particles composed of  dark photons, dark electrons, etc. Such particles could interact in roughly the same way as Standard Model (SM) particles do, but only within their own sector. They could also interact weakly within and across sectors,  the DM sector and  visible matter sector \cite{Yang,Kobzarev,Pavsic,Foot,Akhmedov,Oli11,Oli16}. Particle copies within the dark sectors do not necessarily have the equal masses and couplings as they do in the visible sector. All these thoughts open up a huge number of physics scenarios for DM and its interaction with what we see, the visible matter. 
 
\begin{figure}[ptb]
\begin{center}
\includegraphics[width=7.5cm]{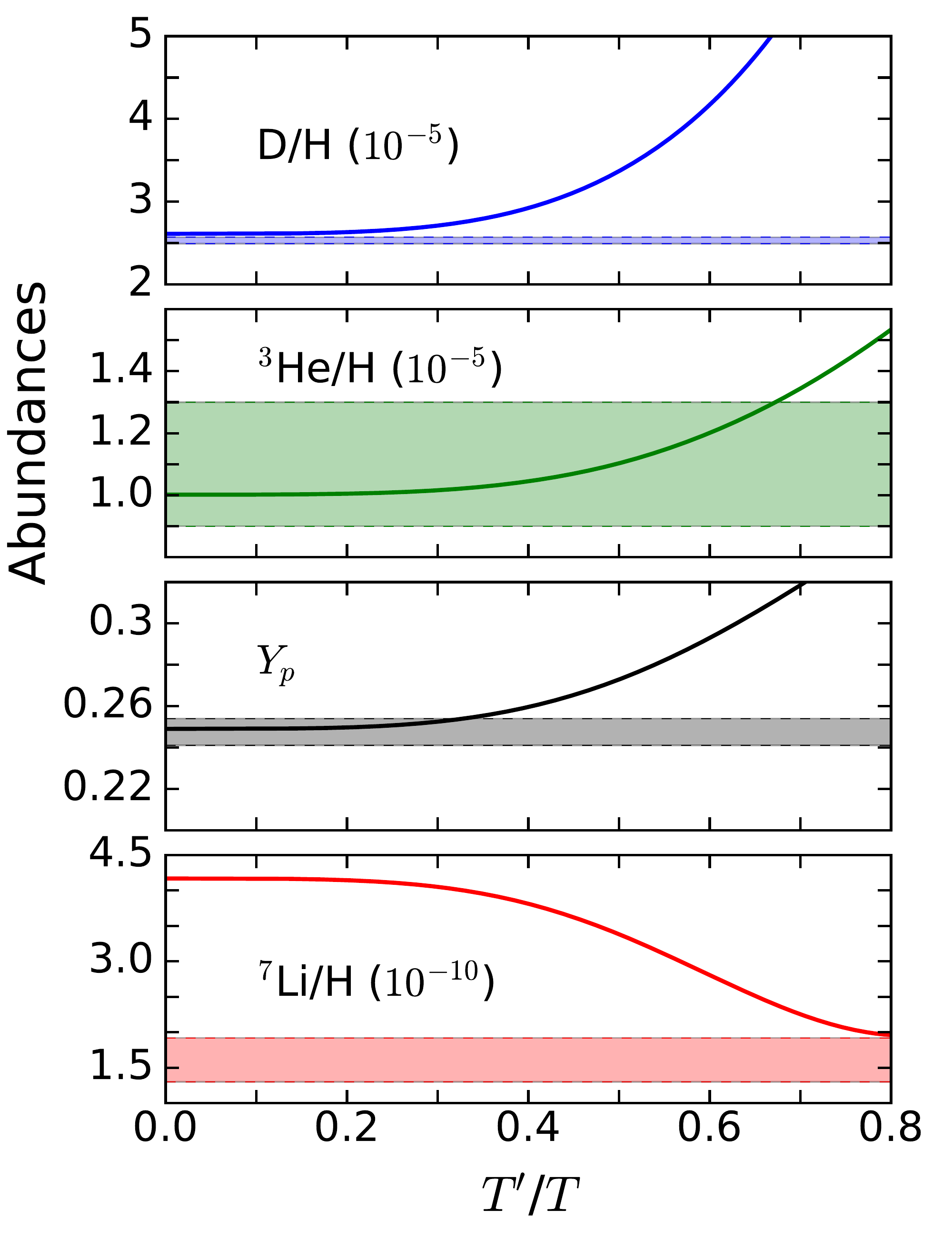}
\end{center}
\caption{BBN model prediction for the relative abundances of D, $^3$He, $^4$He (mass fraction, $Y_p$) and $^7$Li as a function of early universe temperatures ratio between the dark sector temperatures and visible matter temperature, $T'/T$. The calculations were done assuming that the number of dark sectors are  $N_{DM}=5$.  The bands represent uncertainties in the observations \cite{Ber17}. }
\label{dm1}
\end{figure}

Present analysis of astronomical observations lead to a ratio of density parameters $\Omega_{DM} / \Omega_{visible} =4.94 \pm 0.66$. This indicates that DM is about 5 times more massive than visible matter.  This property has been used in Ref. \cite{Oli11,Ber13,Oli16} to study a possible scenario that our universe might consist of of 5 dark parallel universe sectors and not only one single ubiquitous  sector. The motivation for this hypotheses is the assumption that all parallel universes contain nearly the same mass. A Weakly Interacting Massive Gauge Boson (WIMG) was proposed as a way to couple the different dark sectors and the ordinary matter with each dark universe. It was found that a massive WIMG, of the order of  $E\sim 10$ TeV,    does not change the properties of gravity and the SM. It was also shown that this WIMG model is consistent with BBN predictions and CMB observations, but it might have an impact on the lithium abundance.  

The electroweak scale energy, i.e., the temperature of electroweak symmetry breaking, $159.5 \pm 1.5$ GeV, serves as an energy scale for comparisons. Much below this scale, particles in the model can be assumed  to be massless. We group particles into matter/charge fields and a similar structure is used for DM. The generation of the WIMG mass is accomplished by means of a real scalar field, with a very short-range interaction supplied by the WIMG. Our formalism predicts that the number of dark sectors can have an overwhelming role which has been neglected in standard and non-standard BBN models. The obvious reason is that new  degrees of freedom  in the dark sectors can considerably modify the early expansion rate of the universe  \cite{Berezhiani1996} and consequently the light element abundances \cite{Oli11,Ber13,Oli16}.   

The radiation and entropy densities during the BBN epoch are related to the degrees of freedom of particles via
$  \rho(T)=({\pi^{2}}/{30}) \, g_*(T) \,T^4 $ and $ s(T)=({2\pi^{2}}/{45}) \, g_s(T)\, T^3$ with
\begin{eqnarray}
g_*(T)&=&\sum_B g_B \left(\frac{T_{B}}{T}\right)^4 + \frac{7}{8} \sum_F g_F \left(\frac{T_{F}}{T}\right)^4, \ \ \ \  \nonumber \\
g_s(T)&=&\sum_B g_B \left(\frac{T_{B}}{T}\right)^3 + \frac{7}{8} \sum_F g_F \left(\frac{T_{F}}{T}\right)^3, \label{dof}
\end{eqnarray}
with $g_*$ and $g_s$ being the number of degrees of freedom (d.o.f.), and $g_{B(F)}$ the fractions contributed by  bosons (fermions) at the temperatures $T_{B(F)}$, while $T$ is the temperature of the radiation thermal bath.

With minimal assumptions, only two temperatures, $T$ for the ordinary matter sector, and $T'$ for the dark sectors is used. It follows that the
energy $\rho'(T')$ and the entropy $s'(T')$ densities within the dark sectors are also described via Eqs. (\ref{dof}) with the replacements $g_*(T) \rightarrow g'_*(T')$, $g_s(T) \rightarrow g'_s(T')$, and  $T\rightarrow T'$. This implies that $x=(s'/s)^{1/3}\sim T'/T$ emerges from entropy  conservation  in all sectors. With each dark sector having the same total matter as the  visible universe,  it follows that $g_s(T_0) = g_s^\prime(T'_0)$, leading to $x=T'/T$. 

We recall that the Friedman equation is $ H(t)=\dot{a}/a=\sqrt{\left(8\pi/3 c^2\right) \, G_{N} \, \bar{\rho}}$, with $\bar{\rho}$ being the total energy density. Adding to it the matter from a number of dark sectors $N_{DM}$,  the density becomes  $\bar{\rho} = \rho\, + \, N_{DM} \, \rho'$. Thus, the Friedmann equation becomes
$H(t)=1.66 \, \sqrt{\bar{g}_{*}(T)} {T^2}/{M_{Pl}}$, where $\bar{g}_{*}(T) = g_{*} (T) \left( 1+ N_{DM} \, a \, x^4 \right)$, $T(t)$ evolves with time, 
 $M_{Pl}$ denotes the  Planck mass and $a = \left(g'_{*}/g_{*}\right) \left(g_{s}/g'_{s}\right)^{4/3}\sim 1$, for a not too small $T'/T$  \cite{Berezhiani1996}. At  about 1 MeV,  standard BBN assumes $g_{*}(T=1 \ {\rm MeV}) = 10.75$. 
 
 With the  additional dark particles, the number of d.o.f. becomes   $\bar{g}_{*} = g_{*} \left(1 + N_{DM} \, x^4 \right)$. The bounds of $N_{DM}$ and $x$, or $T'/T$ can be studied by comparing the results of BBN calculations including the new d.o.f. with the relative abundances of the light  isotopes D, $^{3}$He, $^{4}$He, and $^{7}$Li.  Our result is displayed in Figure \ref{dm1} with the abundances calculated as a function of $T'/T$ for a fixed number of dark sectors $N_{DM}=5$. The shaded bands displays the uncertainties related to the observed abundances. The observations of primordial elements of D, $^{3}$He, and $^{4}$He  are compatible with early universe temperatures such that we get $T'/T\sim 0.2-0.3$. 

The so-called $^7$Li puzzle cannot be solved by invoking this model  because if one uses $T'/T\sim 1$ it implies that the abundance of $^7$Li becomes compatible with  observations (see Fig. \ref{dm1}, bottom panel), but  then  the abundances of the other elements become  in complete disagreement with observations.  In fact, the d.o.f. factor $\bar{g}_{*}(T)$ depends much more strongly on $T'$ than a variation of $N_{DM}$. For example, if one uses $T'=0.3T$ for cold dark sectors, a large range of values for  $N_{DM}=1-50$ compatible with the D, $^3$He and $^4$He  abundances is found \cite{Ber17}.  

\begin{figure}[ptb]
\begin{center}
\includegraphics[width=8.cm]{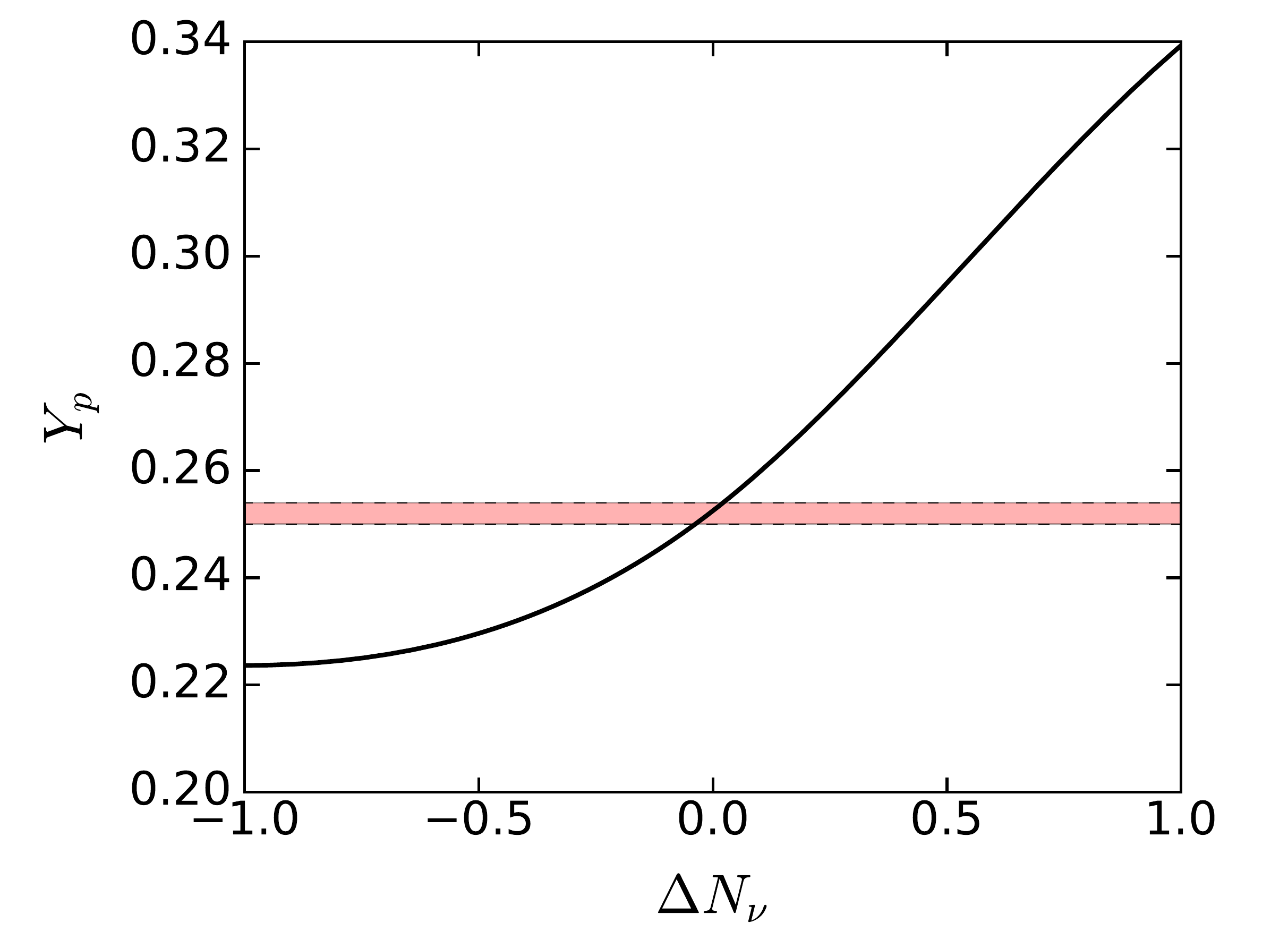}
\end{center}
\caption{ Predictions for the primordial $^4$He mass fraction as a function of extra  neutrino families, $N_\nu =3 + \Delta N_\nu$, with $T'= 0.3T_{BBN}$ and  $N_{DM}=5$. The horizontal band represents the observed mass fraction \cite{Ber17}.}
\label{dm2}
\end{figure}

Neutrinos play an important role for the counting of d.o.f. in the BBN.  Figure \ref{dm2} displays the BBN predictions for the primordial $^4$He mass fraction as a function of extra number of  neutrino families $\Delta N_\nu$ in our parallel universe model, with $T'= 0.3T_{BBN}$ and  $N_{DM}=5$. The horizontal band represents the range of observed mass fraction \cite{Ber17}. The dark sectors model is therefore in accordance with a number of neutrino families $N_\nu =3$. This result remains robust even assuming the existence of the parallel universes of dark matter and the changes in the number of d.o.f. 

We conclude that the existence of multiple cold sectors is open to speculations, and a universe  composed of visible matter and more than one sector of dark matter, e.g. $N_{DM} =5$, and with temperatures of the dark sectors of the order of $T'\sim 0.2-0.3T$ does not influence  the abundances of light elements. The burden of proof for the existence of a rather simple composition of the universe in terms of 5 dark parallel sectors plus a visible matter sector relies simply on us. Our inability to disprove it does not render our claim valid and it just adds to an incredible plethora of speculations about the physics of the early universe, as reported in the literature. As with the physics of neutron stars, BBN is a fertile ground for new theoretical ideas simply because only a few observations can be used as constraints.

\section{The role of statistics for the BBN}

The Maxwell-Boltzmann (MB) distribution  is widely known to reproduce extremely well the distribution of velocities of particles in a thermal bath.   The MB distribution is based on core assumptions adopted by  the Boltzmann-Gibbs statistics. They are: (a) The time between collisions among particles is much longer than their interaction time; (b) the interactions are short-ranged; (c) no correlation exists between the particle velocities;  (d) the collision energy is conserved without transfer to internal degrees of freedom. These are very constraining assumptions, not expected to be always valid even in a system of particles in thermodynamical equilibrium.  In fact, alternatives to the Boltzmann-Gibbs (BG) statistics  exist  and are frequently used in the literature \cite{Ren60,Ts88,GT04}.  

Our group has investigated the impact of using one of the so-called ``non-extensive" statistics to describe the relative velocities of particles during the  BBN  \cite{BFH13}.  Other ideas connected to the statistical aspects of BBN have been explored in the literature (see, e.g., Refs. \cite{KLS94,SFK95}). The non-extensive statistics implies that the entropies of two different microstates $S_A$ and $S_B$ of a system do not add in an extensive way (such as volume). That is, $S \neq S_A + S_B$.  One of the most popular non-extensive statistics was reported in Refs. \cite{Ts88,GT04}. We have calculated the effect on the BBN light element abundances  using the so-called  Tsallis statistics \cite{BFH13}. The Tsallis statistics depends on an additional parameter $q$ representing its departure from the Boltzmann statistics. Nobody knows for sure what the physical meaning of this parameter is and maybe there is none, besides the fact that the Boltzmann statistics is recovered in the limit $q = 1$. But one knows that this statistics has been extremely successful in describing a large number of experimental data not in agreement with the Boltzmann statistics.

\begin{figure}[ptb]
\begin{center}
\includegraphics[width=7.cm]{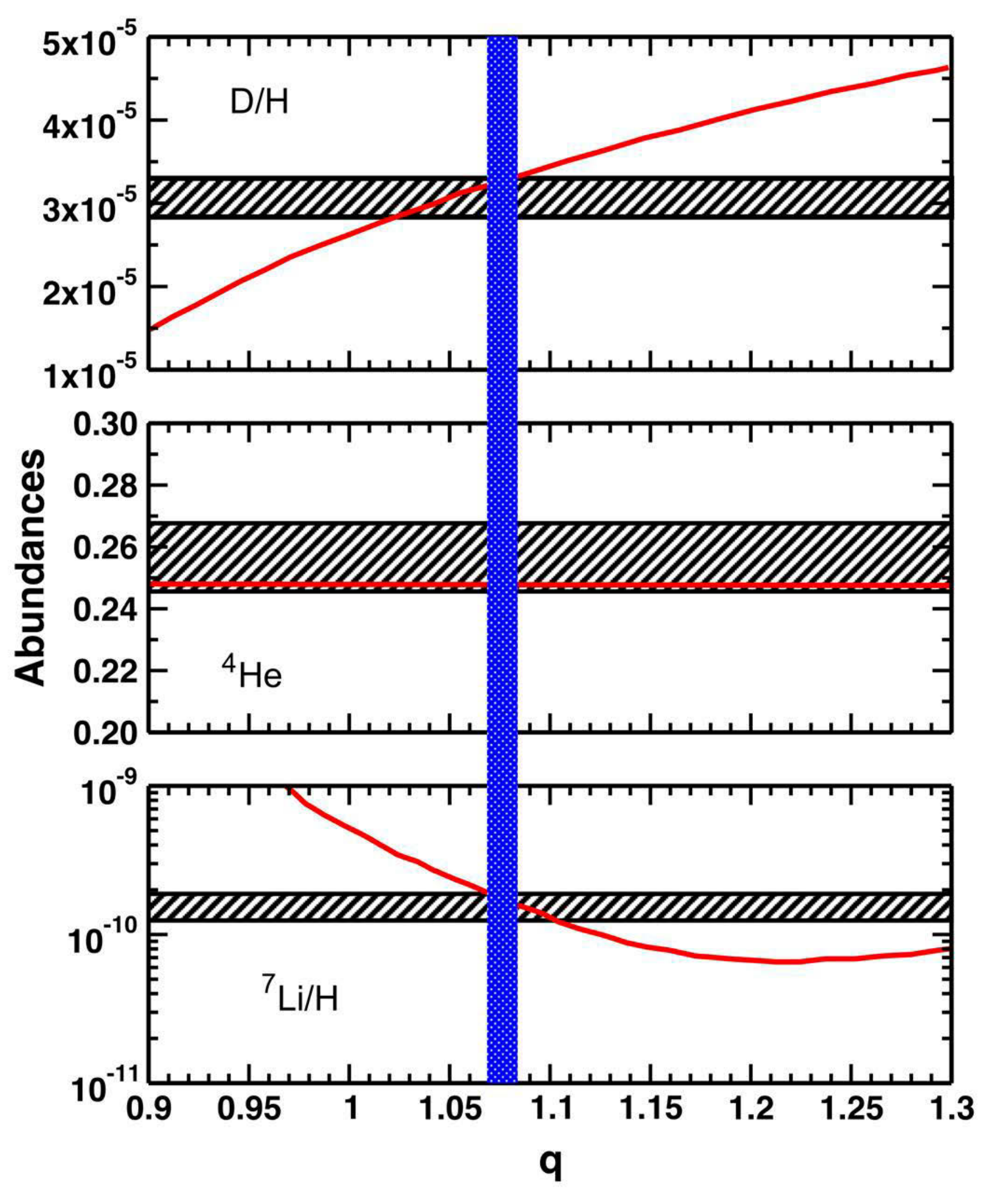}
\end{center}
\caption{Predicted abundances D, $^4$He and $^7$Li (red curves) as a function of the Tsallis parameter $q$  \cite{Hou17}. The observed primordial abundances including 1$\sigma$ uncertainties are indicated by hatched horizontal bands \cite{Aver21,Sbo10,Oliv12}. The vertical (blue) band refers to the parameter $q$ within the interval $1.069<q<1.082$.}
\label{lipuzzle}
\end{figure}

Previous applications of non-extensive statistics, seem to indicate that the non-extensive parameter $q$ does not need to be very different than the Boltzmann value $q=1$ to explain many physics phenomena not directly explainable with $q=1$.  Non-extensive Maxwellian velocity distributions have been used in the past to study stellar nuclear burning  \cite{MQ05,HK08,Deg98,Cor99}.  But in Ref. \cite{BFH13} it was first used for the reaction rate calculations during the  BBN to obtain the  ${^4}$He, D, ${^3}$He, and ${^7}$Li abundances and study its impact on the lithium puzzle. The reaction rates for astrophysics are calculated using the average $\langle \sigma v \rangle$ around the Gamow window, where $\sigma$ is the velocity dependent reaction cross section and $v$ the ions relative velocit (see Eq.\eqref{avemb}). In Ref. \cite{BFH13} a non-extensive MB distribution for $v$ was used in the BBN model. We have included reaction rates for p(n,$\gamma$)d, d(p,$\gamma){^3}$He, d(d,n)${^3}$He, d(d,p)t, ${^3}$He(n,p)t, t(d,n)${^4}$He, ${^3}$He(d,p)${^4}$He, ${^3}$He$(\alpha,\gamma){^7}$Be, t$(\alpha,\gamma){^7}$Li, ${^7}$Be(n,p)${^7}$Li and ${^7}$Li(p,$\alpha){^4}$He. These were fitted to cross sections based on  presently available experimental data \cite{BFH13}.  Our initial conclusion was that if either $q>1$ or $q<1$ the abundances of all elements were somehow affected, but  the $^7$Li abundance always increased   \cite{BFH13}. Thus, we concluded that the lithium puzzle would always  get worse  if extensive statistics is used in the BBN model.

A small, but relevant physics input was neglected in Ref. \cite{BFH13}. This related to the proper modification of the reaction Q-values for backward reaction rates modified with the non-extensive statistics. Such a modification was  considered in Ref.  \cite{Hou17}. It led to amazing results with an impressive agreement with the observed abundances, including that of lithium, with a relatively small departure of the parameter $q$ from unity. The   H, D, $^3$H, $^3$He, and $^4$He abundances remain basically unchanged, but the $^7$Li abundance is appreciably modified, thus solving the $^7$Li puzzle.  The excellent agreement, shown in Fig.  \ref{lipuzzle} for D, $^4$He, and $^7$Li , is obtained using $1.069<q<1.082$. One might argue that this agreement is fortuitous, if one is skeptical of non-extensive statistics. But this result shows that a possible solution to the cosmological lithium problem is due to a fine tuning of the physics involved in the BBN model. 

In Fig.  \ref{lipuzzle} the primordial abundances include 1$\sigma$ uncertainties from observations, indicated with hatched horizontal bands \cite{Aver21,Sbo10,Oliv12}. The vertical (blue) band refers to the value of the parameter $q$ for the  already mentioned interval $1.069<q<1.082$. Ref. \cite{Hou17} was promoted as a research highlight of the American Astronomical Society \cite{AAS17}, attesting the serious concern the Society has with the long standing lithium puzzle. It also displays a certain anxiety in the astronomy community that a reasonable physics explanation is long due, be it in the realm of stellar astration of light elements, cosmic ray spallation processes, or some other physics process. 
It is worthwhile mentioning that the agreement with D/H observations displayed in Fig. \ref{lipuzzle}  would not be as good using the more recent compilation
\begin{equation}
({\rm D/H})_{obs} = (2.55 \pm 0.03) \times 10^{-5},
\end{equation}
based on an weighted average of 11 measurements \cite{Yeh21}. 

The exercises played in Refs. \cite{Oli11,Ber13,Oli16,BFH13,Hou17} are based on ingenious ideas that some times are valid in order to explore a fine tuning of all possible physics processes. It is a trend that new physics ideas flourish when unexpected results emerge from experiments and observations. 

\medskip

{\bf Acknowledgements}

This work was partially supported by the U.S. DOE grant DE- FG02-08ER41533.

\end{document}